\documentclass[cits]{PoS}
\title{Determining QCD Low-Energy Couplings from lattice simulations}
\ShortTitle{Determining QCD Low-Energy Couplings from lattice simulations}
\author{\speaker{Silvia Necco}\thanks{Supported by Marie Curie Fellowship MEIF-CT-2006-025673. Work partially supported by EC Sixth Framework Program under the contract MRTN-CT-2006-035482 (FLAVIAnet), 
Ministerio de Educaci\'on y Ciencia (MEC) under grant FPA2004-
00996 and Generalitat Valenciana under grant GVACOMP2007-156.
Preprint number: IFIC/07-61
}\\
        Instituto de F\'isica Corpuscular, CSIC-Universitat de Val\`encia\\
	Apartado de Correos 22085, E-46071 Valencia, Spain\\
        E-mail: \email{necco@ific.uv.es}}

\abstract{
Different strategies for the computation of QCD low-energy couplings by matching lattice QCD with the chiral effective theory are reviewed. After recalling the main features of the chiral effective theory in the $\epsilon$- and $p$- regimes, the current status of the determination of leading order ($\Sigma$, $F$) and next-to-leading order ($l_i$, $L_i$) low-energy constants is summarised, focusing in particular on recent results obtained with $N_{\rm f}=2$ and $N_{\rm f}=2+1$ simulations.}

\FullConference{The XXV International Symposium on Lattice Field Theory\\
		 July 30-4 August 2007\\
		 Regensburg, Germany}

\usepackage{amsmath,amssymb}
\newcommand{\Dslash}{D\mkern-11.5mu/\,}
\begin{document}
\section{Introduction}
\noindent
QCD dynamics at low momenta can be described in terms of a chiral effective theory, which is  formulated by assuming chiral symmetry and its spontaneous breaking. The effective Lagrangian is expanded as \cite{Weinberg:1978kz,Gasser:1983yg,Gasser:1984gg}
\begin{equation}
\mathcal{L}=\mathcal{L}_{\chi}^{(2)}+ \mathcal{L}_{\chi}^{(4)}+\cdots,
\end{equation}
where 
\begin{eqnarray}
\mathcal{L}_{\chi}^{(2)} & = & \frac{F^2}{4}{\rm Tr} \left[\partial_{\mu}U^\dagger \partial_{\mu}U\right] -\frac{\Sigma}{2}
{\rm Tr} \left[e^{i\theta/N_{\rm f}}\mathcal{M}U+ U^\dagger \mathcal{M}^\dagger e^{-i\theta/N_{\rm f}}          \right],\label{lagr}\\
\mathcal{L}_{\chi}^{(4)} & = & \sum_{i}\mathcal{C}_i\mathcal{O}_i.
\end{eqnarray}
$U\in {\rm SU}(N_{\rm f})$ parametrises the pseudo-Goldstone bosons degrees of freedom; $\Sigma$ and $F$ are the (infinite volume) quark condensate and pseudoscalar decay constant in the chiral limit \footnote{In the following $F$ and $\Sigma$ will refer to the case $N_{\rm f}=2$; for $N_{\rm f}=3$ the notation $F_0$, $\Sigma_0$ will be used.   }; $\mathcal{M}$ is the $N_{\rm f} \times N_{\rm f}$ quark mass matrix. For completeness, the dependence on the vacuum angle $\theta$  is also shown. At NLO, 10 terms appear for $N_{\rm f}=2$, whereas $\mathcal{L}_{\chi}^{(4)}$ contains 12 terms for $N_{\rm f}=3$, with corresponding couplings
\begin{eqnarray}
\{\mathcal{C}_i  \} & \rightarrow & l_{i=1..7},h_{i=1..3}\;\;\;\; [N_{\rm f}=2],\\
\{\mathcal{C}_i  \} & \rightarrow & L_{i=1..10},H_{i=1..2}\;\;\;\; [N_{\rm f}=3].
\end{eqnarray}
A more comprehensive introduction on Chiral Lagrangians has been presented by J. Bijnens at this conference \cite{lat07:Bijnens}. $F$, $\Sigma$, $\{l_i\}$ ($\{L_i\}$) are the so-called \emph{Low-Energy couplings} (LECs): they parametrise the low-energy dynamics, which is not determined by symmetries and can in principle be derived from QCD by means of a non-perturbative method.
Once the LECs are known, chiral perturbation theory becomes a powerful predictive framework for investigating low-energy properties.
Lattice QCD represents the ideal tool to match QCD with the chiral effective theory and to extract LECs in a reliable way;
unquenched simulations are now reaching masses and volumes where such a matching can be performed, and many efforts have been made with the final goal to keep all systematic errors under control.

On a finite volume $V=L^3T$ with $L\gg 1/\Lambda_{\rm QCD}$, different chiral regimes can be distinguished.\\
Approaching the chiral limit by  keeping $M_\pi L\gg 1$ defines the so-called $p$-regime, where the power counting 
in terms of the momentum $p$ and the quark mass $m$ is given by
\begin{equation}
m\sim p^2,\;\;\;\;\; 1/L,\;1/T\sim p.
\end{equation}
In this regime, the chiral effective theory in a finite box looks very much like in infinite volume:
finite-volume effects are exponentially suppressed by factors $\sim\exp(-M_\pi L)$, while mass-effects are dominant.

Alternatively, one can approach the chiral limit by keeping $\mu=m\Sigma V\lesssim O(1)$; this defines the so-called $\epsilon$-regime \cite{Gasser:1986vb,Gasser:1987ah}, where the Compton wavelength associated to the pseudo-Goldstone bosons is larger than the linear extend $L$ of the box, $M_\pi L < 1$ \footnote{The case $M_\pi L < 1$ and $T\gg L$, corresponding to the so-called $\delta$-regime, will not be considered in this discussion.}. 
In this case the power counting is reorganised such that
\begin{equation}
m\sim \epsilon ^4,\;\;\;\;\; 1/L,\;1/T\sim \epsilon.
\end{equation}
It follows that mass effects are suppressed, while volume effects are enhanced and become polynomial in $L^{-2}$.
One of the consequences of the rearrangement is that, at a given order in the effective theory, less LECs appear with respect to the $p$-regime: the NLO predictions are less contaminated by higher order effects, making the $\epsilon$-regime particularly advantageous and clean to compute the leading order couplings $F$ and $\Sigma$.
Furthermore, in the $\epsilon$-regime topology plays a relevant r\^ole \cite{Leutwyler:1992yt}. Observables can be defined at fixed value of the topological charge, and the dependence on this charge should also be well reproduced by the effective theory. Therefore topology provides a new variable in this regime, in addition to the mass and the volume.
In the infinite volume limit, this dependence is expected to vanish. 

The possibility of matching the chiral effective theory with QCD in different regimes and by means of different observables opens the chance for several independent determinations of LECs -with different higher order corrections and systematic errors - from which a precise and first-principles determination of the LECs can be achieved in the near future.

\section{Lattice determinations of LECs in the $p$-regime}
\subsection{$N_{\rm f}=2$}
\noindent
In QCD with two flavours, the chiral expansion of the pion mass and pion decay constant at NLO in the isospin limit is given by \cite{Gasser:1983yg}
\begin{eqnarray}
M_\pi^2 & = & M^2+\frac{M^4}{32\pi^2 F^2}\ln\left(\frac{M^2}{\Lambda_3^2}  \right)+\cdots,\\
F_\pi   & = & F-\frac{M^2}{16\pi^2F}\ln\left(\frac{M^2}{\Lambda_4^2}  \right)+\cdots,
\end{eqnarray}
with 
\begin{equation}
M^2=\frac{2m\Sigma}{F^2}.
\end{equation}
The scale-independent LECs $\overline{l}_3,\overline{l}_4$ are defined as
\begin{equation}
\overline{l}_{3,4}\equiv\ln \left(\Lambda_{3,4}^2/M^2\right)_{M=139.6\;{\rm MeV}}.
\end{equation}
The values of these parameters extracted from phenomenological analysis are \cite{Gasser:1983yg,Bijnens:1996wm,Colangelo:2001df,Colangelo:2003hf}
\begin{equation}
F= 86.2(5)\; {\rm MeV},\;\;\;\;\overline{l}_3=2.9(2.4),\;\;\;\;\overline{l}_4=4.4(2).
\end{equation}
The full NNLO expression for those quantities is also known (see \cite{lat07:Bijnens}). \\
In many recent lattice studies with $N_{\rm f}=2$ dynamical fermions the quark mass dependence of $M_\pi$ and $F_\pi$ has been studied in order to extract $F$, $\Sigma$, $\overline{l}_3$, $\overline{l}_4$. The main characteristics of the different computations are summarised in table \ref{preg_par}, while the results are listed in table \ref{preg_res}; notice that the quark condensate is given in the $\overline{\rm MS}$ scheme at energy scale 2 GeV.\\
The first attempt to estimate directly $\overline{l}_{3,4}$ from unquenched lattice calculations has been performed in \cite{DelDebbio:2006cn}, where two-flavour QCD with Wilson quarks (with and without non-perturbative on-shell $O(a)$ improvement) has been simulated using DD-HMC algorithm \cite{Luscher:2003qa,Luscher:2005rx}.
The authors pointed out that the extraction of $\overline{l}_4$ is problematic, since the obtained results are not stable by adding higher order terms in the NLO fit of the pseudoscalar decay constant. The error quoted in table \ref{preg_res} for  $\overline{l}_3$ is statistical only.

ETM collaboration is carrying on an extensive project with $N_{\rm f}=2$ Wilson twisted mass quarks at maximal twist.
 First results with light quarks have been presented in \cite{Boucaud:2007uk}.
Updated results have been presented at this conference \cite{lat07:Urbach,lat07:Herdoiza} and are reported in table \ref{preg_res}; the data have been corrected for finite-size effects using the extended L\"uscher formula \cite{Colangelo:2005gd}.
The quoted errors are statistical, $O(a^2)$ effects and finite-size effects respectively. Notice that in this study the scale has been fixed through the physical values of $M_\pi$ and $F_\pi$. See \cite{lat07:Urbach,lat07:Herdoiza} for the complete set of ensembles produced by the collaboration.

QCDSF and UKQCD are performing $N_{\rm f}=2$ simulations using Wilson fermions with non-perturbative on-shell $O(a)$ improvement. Several lattice spacings and volumes have been simulated (see \cite{lat07:Schierholz} for the full list). 
The data are corrected for  $O(a^2)$ effects and for finite-size effects estimated from \cite{Colangelo:2005gd}. The scale in this case has been determined from the axial coupling $g_A$.

Finally, JLQCD presented at this conference the first results obtained with overlap Dirac operator and Iwasaki gauge action \cite{lat07:Matsufuru, lat07:Noaki}. 
Two kinds of finite volume corrections have been taken into account: the usual finite volume effect computed according to \cite{Colangelo:2005gd}, and the correction due to fixed topology, according to \cite{Aoki:2007ka}. Large differences have been observed by adding 2-loop terms in the chiral fit, with respect to a NLO fit;
two different NNLO chiral fits have been tried, and the relative difference considered as systematic uncertainty (second error in table \ref{preg_res}). Here the scale has been fixed by the condition $r_0=0.49$ fm \cite{Sommer:1993ce}.

All these results fairly agree among themselves; although systematic errors need to be better quantified, 
we expect them to be under control in the next years, so that 
the path towards a reliable computation of LECs of two-flavour chiral theory looks very promising. 

The phenomenological impact of the lattice computations of $\overline{l}_3,\overline{l}_4$ has been discussed recently by H. Leutwyler \cite{Leutwyler:2007ae} ; in particular he considered the $S$-wave pion scattering lengths $a_0^I\;(I=0,2)$, which can be related to  $\overline{l}_3,\overline{l}_4$ by means of low-energy theorems \cite{Colangelo:2001df}. 
\begin{table}
\begin{center}
\begin{tabular}{|l|l|l|l|l|l|}
\hline
&&&& & $M_{\pi,\rm{min}}$ \\
{\emph{Authors}}     & {\emph{Dirac op.}} & {\emph{gauge act.}} & $a$ (fm) & $T/a,\;L/a$ &  (MeV) \\
\hline
Del Debbio et al \protect\cite{DelDebbio:2006cn} &  Wilson            &  Wilson             & 0.0717(15)   & 32, 24 &   403\\
                                                 &              &               & 0.0521(7)    & 64, 32 & 381 \\
                                                 & Wilson $O(a)$ impr.&  Wilson             & 0.0784(10)    & 48, 24 & 377 \\ 
\hline
ETM \protect\cite{lat07:Urbach,lat07:Herdoiza}   & Wilson Twisted Mass &  Sym. tree          & 0.0858(5)    & 48, 24 & 300      \\
                                                 & &          & 0.0858(5)    & 64, 32 & 300      \\
                                                 & &         & 0.0657(11)   & 48, 24 & 420      \\
                                                & &         & 0.0657(11)   & 64, 32 & 300      \\
\hline 
QCDSF/UKQCD  \protect\cite{lat07:Schierholz}     &  Wilson $O(a)$ impr. & Wilson         & $\simeq$ 0.08   & 48, 24   & 450                  \\
                                                 &  &        & $\simeq$ 0.08   & 64, 32   & 340 \\   
                                                 & &       & $\simeq$ 0.07   & 64, 32   & 440 \\   
\hline
JLQCD \protect\cite{lat07:Matsufuru,lat07:Noaki}       &  Overlap & Iwasaki         & $\simeq$ 0.12   & 32, 16   & 290 \\  
\hline
\end{tabular}
\caption{Parameters of recent $N_{\rm f}=2$ simulations in the $p$-regime with light quarks. The second column refers to the Dirac operator, the third to the gauge action adopted. In the sixth column, $M_{\pi,{\rm min}}$ indicates the lightest pion mass reached in the simulations.}   \label{preg_par}
\end{center}
\end{table}
\begin{table}
\begin{center}
\begin{tabular}{|l|l|l|l|l|}
\hline
&&&& \\[-0.3cm]
{\emph{Authors}}     & $F$  &   $\Sigma^{\overline{\rm MS}}$(2 GeV) &  $\overline{l}_3$ &  $\overline{l}_4$  \\
     & (MeV)  &     $({\rm MeV})^3$        &               & \\
\hline
Del Debbio et al \protect\cite{DelDebbio:2006cn} &                       &              &  3.0(5) &   \\
\hline
ETM   \protect\cite{lat07:Urbach,lat07:Herdoiza}            & 85.98(7)(21)(35) &  $266(6)(0)(6)$  & 3.44(8)(26)(6) & 4.61(4)(3)(7) \\
\hline
QCDSF/UKQCD  \protect\cite{lat07:Schierholz}       & 79(5)                 &  273(12)      &  3.49(12) &  4.69(14)\\
\hline
JLQCD   \protect\cite{lat07:Matsufuru,lat07:Noaki}        & 78(3)(1)        &   $242(6)(6)$   & 2.9(4)(2.6) & 4.3(5)(2)\\
\hline
\end{tabular}
\caption{Summary of LECs obtained from $N_{\rm f}=2$ simulations.} \label{preg_res}
\end{center}
\end{table}

\subsection{$N_{\rm f}=2+1$}
\noindent
Lattice results obtained with three dynamical flavours can be matched with $N_{\rm f}=3$ chiral perturbation theory in order to extract $\Sigma_0,F_0,\{L_i\}$.
In particular, in the three-flavour chiral effective theory, the quark-mass dependence of pseudoscalar decay constants at NLO contains the LECs $L_4$ and $L_5$,
while pseudoscalar masses are sensitive to the combinations $(2L_8-L_5)$ and $(2L_6-L_4)$ \cite{Gasser:1984gg}. Moreover, the same LECs can be extracted from a partially quenched setup \cite{Sharpe:2000bc,Bijnens:2006jv} (see also \cite{Sharpe:2006pu}). \\
The current status of phenomenological determinations of the $L_i$ has been summarised by Bijnens in this conference \cite{lat07:Bijnens}; from a NLO fit one infers 
\begin{equation}
L_4\equiv 0;\;\;\;\; L_6\equiv 0;\;\;\;\; 10^3\cdot L_5=1.46;\;\;\;\; 10^3\cdot L_8=1.00.
\end{equation}
The parameters related to recent  $N_{\rm f}=2+1$ lattice simulations are summarised in the top part of table \ref{preg_n3_res}; the corresponding results 
for the renormalised $(2L_8-L_5)$, $(2L_6-L_4)$, $L_4$, $L_5$ evaluated at the scale $M_{\rho}=770$ MeV are collected in the bottom part.

The MILC collaboration presented the first results obtained with staggered fermions and fourth root prescription in 2004
\cite{Aubin:2004fs}; several updates followed (see for instance the contribution at the 2006 lattice conference \cite{Bernard:2006wx}) and new preliminary results were presented at this conference \cite{lat07:Bernard}. 
The chiral fits are performed using (partially quenched) rooted staggered chiral perturbation \cite{Lee:1999zxa} including analytic NNLO and NNNLO terms. By setting the scale through $F_\pi$ they obtain for the LO constants
\begin{eqnarray}
F_\pi/F_2 & = & 1.052(2)\textstyle{\binom{+6}{-3}},\;\;\;\;\;\;\;\Sigma_2^{\overline{\rm MS}}(2\;{\rm GeV}) =[278(1)\textstyle{\binom{+2}{-3}}(5)\; {\rm{MeV}}]^3, \\
F_\pi/F_0 & = & 1.21(5)\textstyle{\binom{+13}{-3}},\;\;\;\;\;\;\; \Sigma_0^{\overline{\rm MS}}(2\;{\rm GeV})= [242(9)\textstyle{\binom{+5}{-17}}(4)\;{\rm{MeV}}]^3,                 \nonumber
\end{eqnarray}
where $F_0,\Sigma_0$  refer to the 3 flavour chiral limit, while $F_2,\Sigma_2$ are obtained by sending $m_u,m_d\rightarrow 0$ and $m_s\rightarrow m_{s,\rm phys}$. The quoted errors are statistical and lattice-systematic; for the condensates there is an additional uncertainty coming from perturbative renormalisation.
 
RBC and UKQCD collaborations presented this year first results for light meson masses and pseudoscalar decay constants computed with Domain Wall Dirac operator and Iwasaki gauge action \cite{Allton:2007hx}; in this first work, with $L\simeq 1.8$ fm and pseudoscalar masses up to 400 MeV, they observed that a simultaneous NLO fit of pseudoscalar meson masses and decay constants fails. 
New results have been presented at this conference, with a larger volume and lighter quark masses \cite{lat07:Boyle,lat07:Lin} (in particular, $M_{PS,{\rm min}}\simeq 330$ MeV for the full case and  $M_{PS,{\rm min}}\simeq 250$ MeV in the partially quenched setup).
By observing that three-flavour fits are not safely applicable in the mass range up to the kaon mass, they also performed partially quenched ${\rm SU}(2)\times {\rm SU}(2)$ fits, where only two quarks are treated as light. To compare the two fits, they derived from both cases the constants
$\overline{l}_3,\overline{l}_4$ using NLO matching  between two- and three-flavour LECs \cite{Gasser:1984gg}, finding compatible results within the errors.

Finally, PACS-CS collaboration is undertaking $N_{\rm f}=2+1$ simulations with $O(a)$ non- perturbatively improved Wilson quarks and Iwasaki gauge action, adopting DD-HMC algorithm; preliminary results were presented at this conference \cite{lat07:Kuramashi,lat07:Ukita}. 

Given the poor knowledge of systematic errors, 
a precise comparison of the lattice results at this point is maybe premature.

\begin{table}
\begin{center}
\begin{tabular}{|l |c |c |c|c|c|}
\hline
{\emph{Authors}} & {\emph{Dirac op.}}  & {\emph{gauge action}} & $a$ (fm) & $L$ (fm) & $M_{PS, {\rm min}}$ (MeV)\\
\hline
MILC\protect\cite{lat07:Bernard} & impr. staggered & Sym. 1 loop     & 0.06-0.15 & 2.4-3.4 & 200   \\
\hline
RBC-UKQCD\protect\cite{lat07:Lin}  & Domain Wall & Iwasaki & 0.11  & 1.8-2.6  & 330 \\
\hline
PACS-CS\protect\cite{lat07:Ukita} & Wilson $O(a)$ impr. & Iwasaki & 0.09 & 2.9   &  210 \\
\hline
\end{tabular}

\vspace{0.3cm}
\begin{tabular}{|l |c |c |c|c|}
\hline 
{\emph{Authors}} &   $(2L_8-L_5)\cdot 10^3$  &    $(2L_6-L_4)\cdot 10^3$  & $ L_4\cdot 10^3$  &  $ L_5\cdot 10^3$  \\
\hline
 MILC\protect\cite{lat07:Bernard} & $0.3(1)(1)$ &  $0.3(1){\textstyle{\binom{+2}{-3}}}$ &  $0.1(3){\textstyle{\binom{+3}{-1}}}$   & $1.4(2){\textstyle{\binom{+2}{-1}}}$ \\
\hline
 RBC-UKQCD\protect\cite{lat07:Lin} &  0.247(45)  & -0.002(42) &  0.136(80) &   0.862(99)\\
\hline
 PACS-CS\protect\cite{lat07:Ukita} & -0.23(5)  &  0.10(4) & -0.02(11) & 1.47(13) \\
\hline
\end{tabular}
\caption{On the top: parameters of recent $N_{\rm f}=2+1$ simulations in the $p$-regime. On the bottom: results for
the LECs evaluated at the scale $M_\rho=770$ MeV.}\label{preg_n3_res}
\end{center}
\end{table}

\section{Determinations of LO LECs in the $\epsilon$-regime}  
\noindent
Lattice simulations in the $\epsilon$-regime require small quark masses, and in general the preservation of chiral symmetry at finite lattice spacing would be highly desirable: the spectral gap of the Dirac operator is then bounded from below and this ensures stability of the dynamical simulations. However it is a matter of fact that Ginsparg-Wilson fermions are still very demanding from the numerical point of view, since the construction of the corresponding Dirac operator is very costly with respect to other regularisations.\\
On the other hand, for the Wilson Dirac operator the spectral gap is not guaranteed to have a lower bound and recent studies established an empirical stability condition  \cite{DelDebbio:2005qa}:
\begin{equation}
m\ge m_{\rm min};\;\;\;\;\;m_{\rm min}\propto \frac{a}{\sqrt{V}}.
\end{equation}
In order to reach the region $m\Sigma V\le 1$, still unrealistically large (with the present computing resources) lattice extents would be needed
\footnote{For $O(a)$-improved Wilson fermions there are first indications that the width of the spectral gap distribution has a different scaling \cite{DelDebbio:2007pz}, although definitive conclusions are still missing.}.\\
The spectral gap of the twisted mass Dirac operator is also bounded from below:
an attempt to investigate the $\epsilon$-regime with this discretisation has been presented at this conference \cite{lat07:Shindler}.

Another reason that makes the $\epsilon$-regime particularly challenging is that in the region where  $m\Sigma V \sim O(1)$ one expects large fluctuations associated to low-modes wavefunctions, which can induce large uncertainties on the observables.
These can be however substantially reduced by adopting the so-called \emph{low-mode averaging} technique \cite{DeGrand:2004qw,Giusti:2004yp}.

At leading order in the $\epsilon$-expansion, the partition function of the effective theory reads 
\begin{equation}
\mathcal{Z}=\int_{\rm SU(N)} dU_0\exp\left[\frac{\Sigma V}{2}{\rm Tr} \left(e^{i\theta/N}
\mathcal{M}U_0+ e^{-i\theta/N}U_0^\dagger \mathcal{M}^\dagger  \right)   \right],
\end{equation}
where $U_0$ represents the collective constant field associated to the zero modes.
The partition function at fixed topology $\nu$ is obtained by Fourier-transforming in $\theta$:
\begin{equation}
\mathcal{Z}_\nu =\int_{\rm U(N)}dU_0 \left({\rm det}U_0  \right)^\nu \exp\left[\frac{\Sigma V}{2}{\rm Tr} \left(\mathcal{M}U_0+ U_0^\dagger \mathcal{M}^\dagger  \right)   \right].
\end{equation}
\subsection{Quark condensate from finite-size scaling}
\begin{figure}
\includegraphics[width=7.3cm]{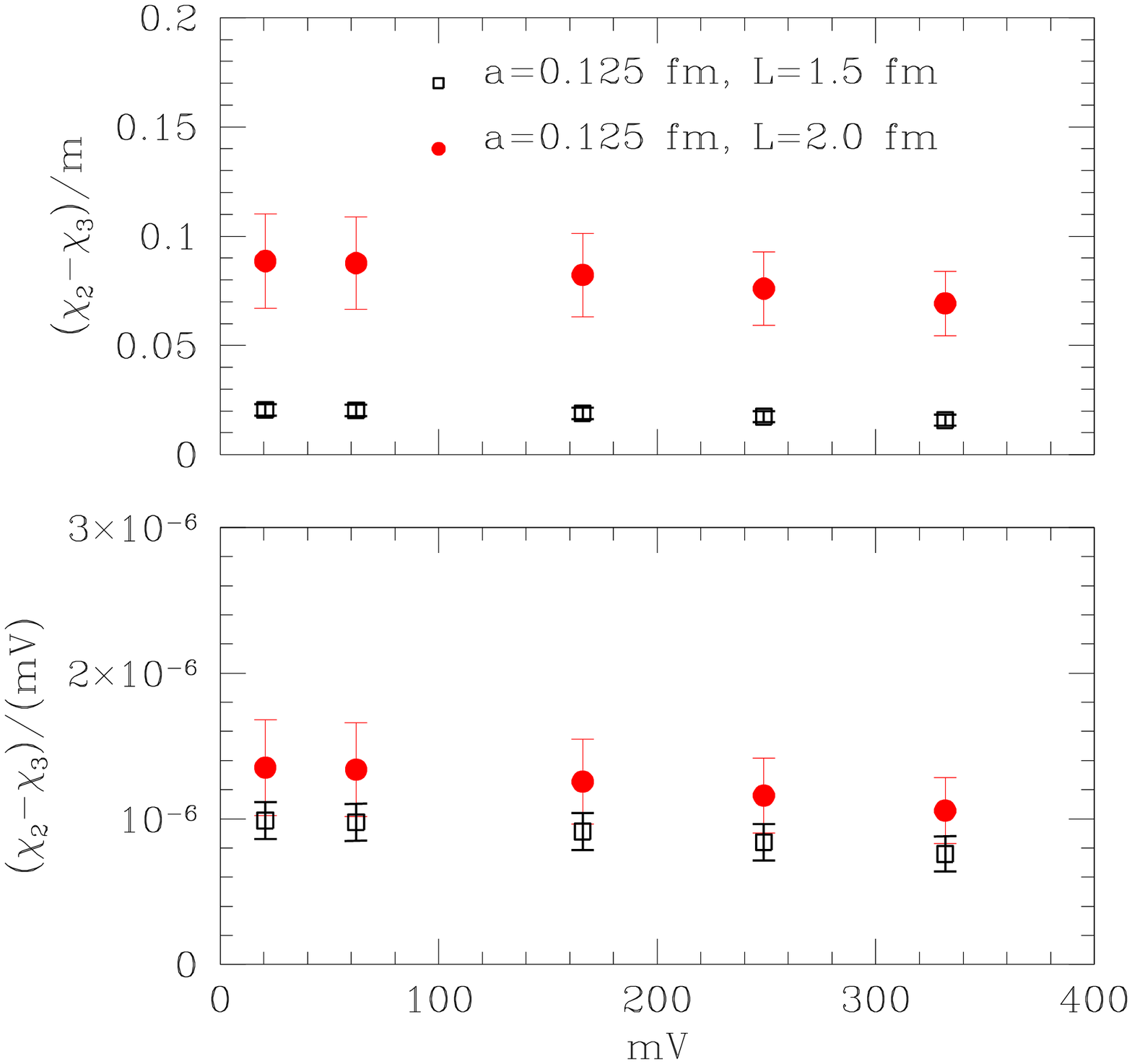}
\includegraphics[width=7.0cm]{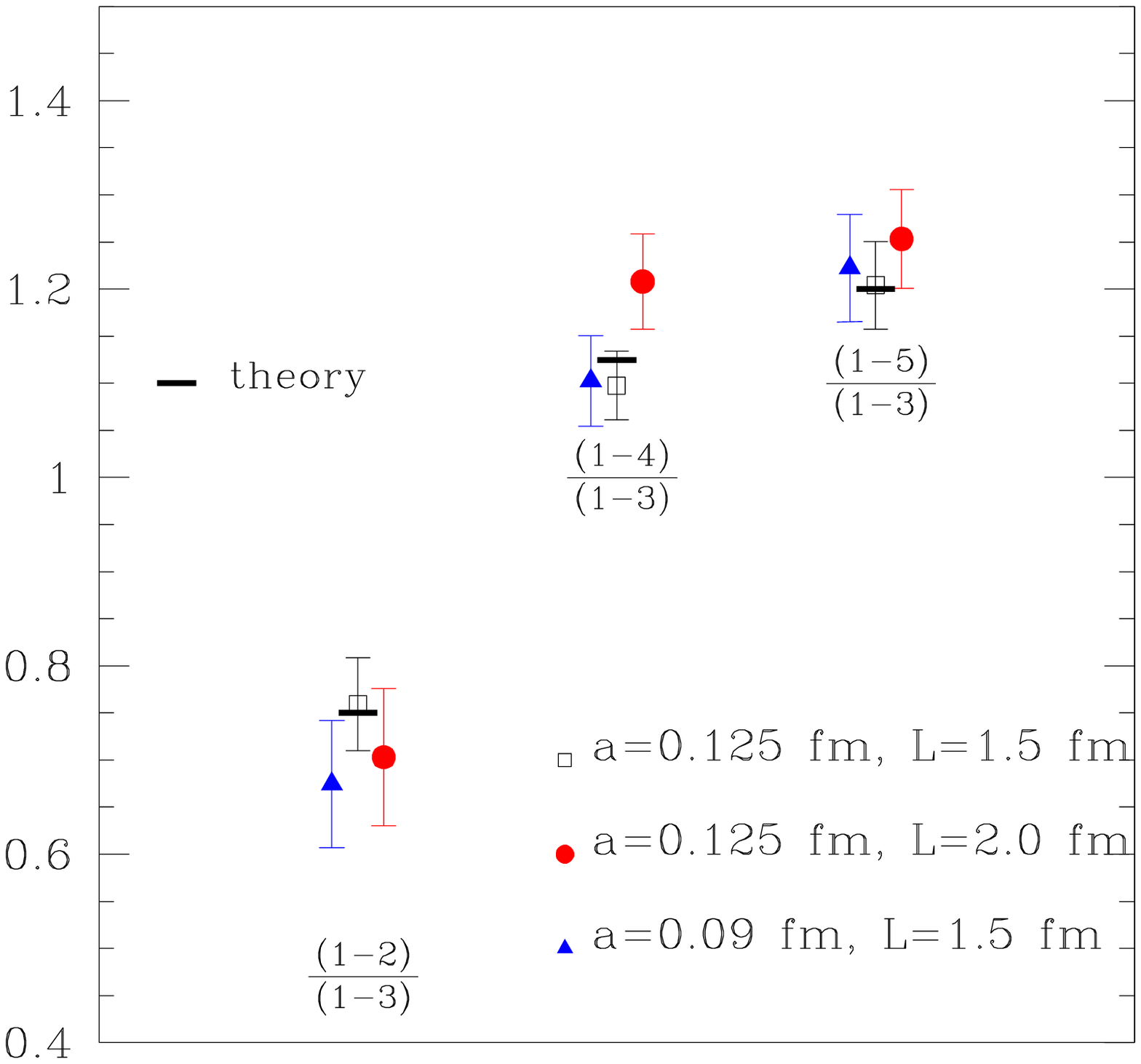}
\vspace{-0.5cm}
\caption{On the left: finite-size scaling of differences $(\chi_2 -\chi_3)$, computed at two different symmetric volumes $V=(1.5\;{\rm fm})^4$ (black empty squares) and $V=(2.0\;{\rm fm})^4$ (red filled points) \protect\cite{Giusticond}. 
The data refer to the quenched approximation, with overlap Dirac operator.
On the right: the ratios $(\chi_{\nu_1}-\chi_{\nu_2})/(\chi_{\nu_3}-\chi_{\nu_4})$ for three different lattices considered in \protect\cite{Giusticond} 
compared with theoretical predictions of eq. \protect\ref{eq:sumru}.}\label{cond_scaling}
\end{figure}

\noindent
We consider for simplicity a quark mass matrix proportional to the identity, $\mathcal{M}=m\mathbf{1}_{N_{\rm f}}$, $N_{\rm f}\ge 2$, and introduce the dimensionless variable $\mu=m\Sigma V$; the quark condensate for $\mu\ll 1$ behaves like \cite{Leutwyler:1992yt}
\begin{equation}
\Sigma(\mu)\equiv\frac{\Sigma}{N_{\rm f}}\frac{\partial}{\partial\mu}\ln\mathcal{Z}\sim \Sigma \mu.
\end{equation}
As expected, $\Sigma(\mu)$ vanishes in the chiral limit since spontaneous symmetry breaking does not occur at finite volume. Analogously, at fixed topology one has \cite{Leutwyler:1992yt}
\begin{eqnarray}
\Sigma_\nu(\mu) & \equiv & \frac{\Sigma}{N_{\rm f}}\frac{\partial}{\partial\mu}\ln\mathcal{Z}_\nu= \frac{\Sigma \nu}{\mu}+\tilde\chi_\nu \label{cond_nu},\\
{\rm{with}}\;\;\;\tilde\chi_\nu & = & \frac{\Sigma}{2(N_{\rm f}+\nu) }\mu+...\label{tildechi},
\end{eqnarray}
where the infrared divergence proportional to $1/m$ is due to zero-modes contribution. From these expressions it is clear that, even if spontaneous symmetry breaking does not occur in a finite volume, the formation of a quark condensate $\Sigma$ leaves signs in a finite box and the corresponding finite-size scaling is predicted by the chiral effective theory. \\
In particular, $\tilde\chi_\nu$ is explicitly known up to NLO. The $O(\epsilon^2)$ contributions are entirely given by one-loop corrections to the chiral condensate \cite{Gasser:1987ah} 
\begin{equation}
\Sigma_{\rm eff}(V)=\Sigma\left[1+\frac{N_{\rm f}^2-1}{N_{\rm f}}\frac{\beta_1}{F^2\hat{L}^2}\right],
\end{equation}
where $\beta_1$ is a known universal shape coefficient \cite{Neuberger:1987fd,Neuberger:1987zz,Hasenfratz:1989pk}  and $\hat{L}=V^{1/4}$. 
The important point is that NLO corrections still contain only LO LECs, a typical fact which occur in the $\epsilon$-regime, as already pointed out in the introduction. \\
From the QCD side, given a lattice regularisation which satisfies the Ginsparg-Wilson relation, the quark condensate at fixed topology can be expressed as
\begin{equation}
-\frac{\langle\overline{\psi}\psi  \rangle_\nu}{N_{\rm f}}=2m\int_0^\infty d\lambda \frac{\rho_\nu(\lambda)}{m^2+\lambda^2}= 
\frac{\nu}{Vm} +\chi_\nu,
\end{equation}
where $\rho_\nu(\lambda)$ is the spectral density of the Dirac operator and 
$1/m$ divergence is exactly the one founded also in the effective theory. Once renormalised with the logarithmic-divergent renormalisation constant $Z_S$, $\chi_\nu$ can be matched with $\tilde\chi_\nu$ of eq. \ref{tildechi}. Notice that at finite mass $\chi_\nu$ contains also additive ultraviolet divergences which must be removed in order to perform a finite-size scaling study. Being topology-independent, a convenient solution for this purpose is to consider the combinations $(\chi_{\nu_1}-\chi_{\nu_2})$, which have an unambiguous continuum limit at finite quark masses. 
Several quenched studies have been performed in order to extract $\Sigma$ from a finite-size scaling study 
\footnote{In the quenched limit $\Sigma_{\rm eff}(V)$ diverges logarithmically with $\hat{L}$ and hence can be defined only at a finite volume.} \cite{Hernandez:1999cu,DeGrand:2001ie,Hasenfratz:2002rp,Giusticond}.
The left side of fig. \ref{cond_scaling} shows the quantity  $(\chi_2 -\chi_3)$ computed in \cite{Giusticond} for two symmetric volumes  $V=(1.5\;{\rm fm})^4$ and $V=(2.0\;{\rm fm})^4$, as a function of $(mV)$. The fact that  $(\chi_2 -\chi_3)/(mV)$ does not depend on the volume within the statistical errors indicates that LO scaling is verified, and there is no sensitivity to NLO corrections. These data have been obtained using low-mode averaging technique. \\
The chiral effective theory gives non-trivial parameter-free predictions in the chiral limit, also known as Leutwyler-Smilga sum rules \cite{Leutwyler:1992yt}
; for instance from eq. \ref{tildechi} one obtains 
\begin{equation}\label{eq:sumru}
\frac{\tilde{\chi}_{\nu_1}(\mu)-\tilde{\chi}_{\nu_2}(\mu) }{\tilde{\chi}_{\nu_3}(\mu)-\tilde{\chi}_{\nu_3}(\mu) }\bigg|_{\mu=0}=\frac{(\nu_1-\nu_2)(\nu_3+N_{\rm f})(\nu_4+N_{\rm f})}{(\nu_3-\nu_4)(\nu_1+N_{\rm f})(\nu_2+N_{\rm f})}.
\end{equation}
A comparison with the lattice data \cite{Giusticond} in the quenched approximation is shown in the right part of fig. \ref{cond_scaling}, for several values of the topological indices and for three different lattices.
The quark mass dependence of the ratios is weak, and the data refer to a very light quark mass corresponding to $\mu\simeq 0.07$. 
The good agreement supports the fact that the topology dependence of those ratios is well reproduced by the (quenched) chiral effective theory.
\subsection{Meson correlators}
\noindent
The pseudoscalar decay constant can be extracted by matching current correlators in the $\epsilon$-regime with the NLO predictions. For example, for the left-handed current correlator at fixed topology
\begin{equation}
C^{ab}(x_0)=Z_J^2\sum_{\vec{x}}\langle J_L^a(x)J_L^b(0)\rangle_{\nu},
\end{equation}
one obtains the following prediction at NLO \cite{Hansen:1990un,Hansen:1990yg,Damgaard:2002qe,Hernandez:2002ds, Hernandez:2006kz}:
\begin{eqnarray}
C^{ab}(x_0) & = & {\rm Tr}\left[T^aT^b \right]\left\{\frac{F^2}{2T}+\frac{N_{\rm f}}{2T} 
\left(\frac{\beta_1}{\sqrt{V}}-\frac{T^2k_{00}}{V}   \right) +
\frac{\mu\Sigma_{\nu}(\mu)}{L^3} h_1\left(\frac{x_0}{T}   \right)  \right\},\\
h_1\left(\frac{x_0}{T}  \right) & = & \frac{1}{2}\left[\left(\left|\frac{x_0}{T}   \right|-\frac{1}{2}\right)^2 -\frac{1}{12}\right].
\end{eqnarray}
$T^a$ are the ${\rm SU}(N_{\rm f})$ generators, $Z_J$ is the renormalisation constant associated to the left current,
$\beta_1$ and $k_{00}$ depend on the geometry and $\Sigma_{\nu}(\mu)$ is defined in eq. \ref{cond_nu}. Also in this case, up to NLO only the LECs $\Sigma$ and $F$ appear. 
Current correlators in the $\epsilon$-regime have been investigated in many quenched studies \cite{Bietenholz:2003bj,Giusti:2004yp,Fukaya:2005yg} using Ginsparg-Wilson fermions. In \cite{Giusti:2004yp} the $J_L J_L$ correlator has been computed both in the $\epsilon$- and $p$-regime: here it has been shown that the quenched value of $F$ obtained in the $p$-regime after a chiral extrapolation is in agreement with the one extracted from the $\epsilon$-regime.

\begin{figure}
\includegraphics[width=8.3cm]{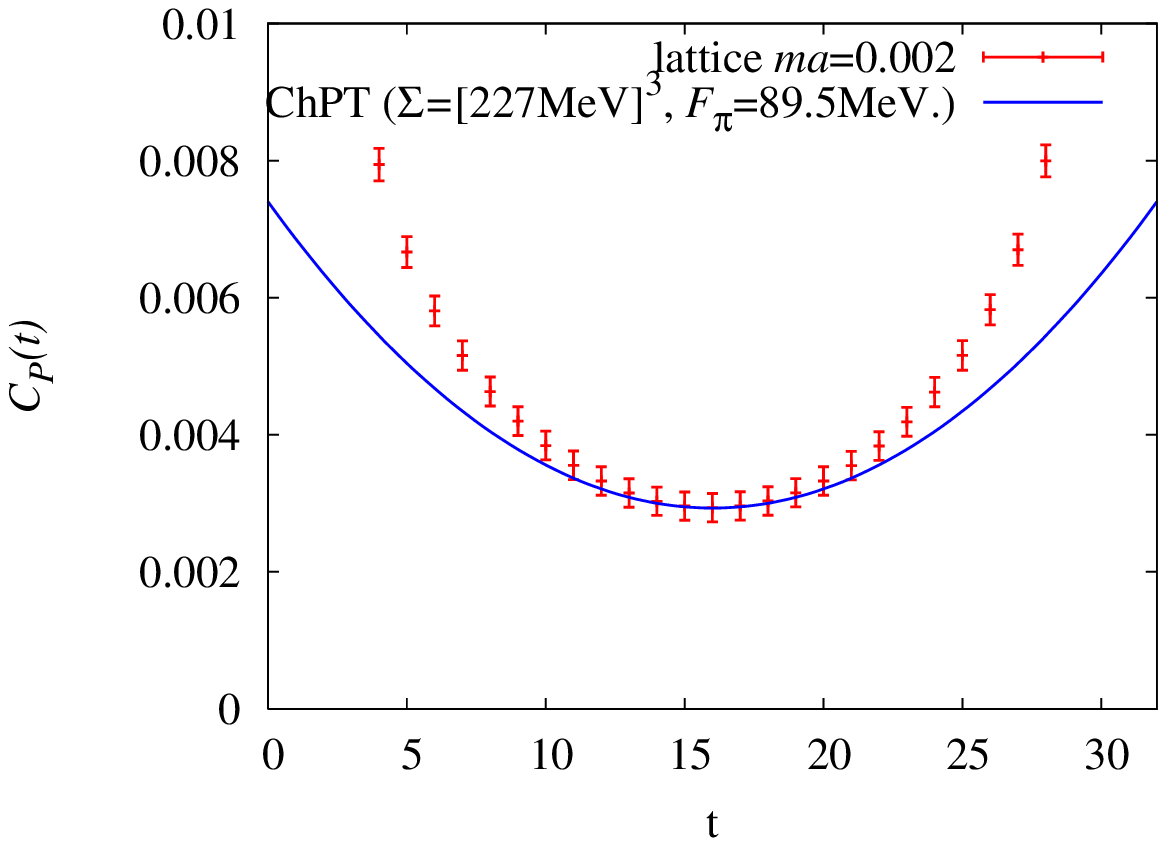}
\includegraphics[width=8.3cm]{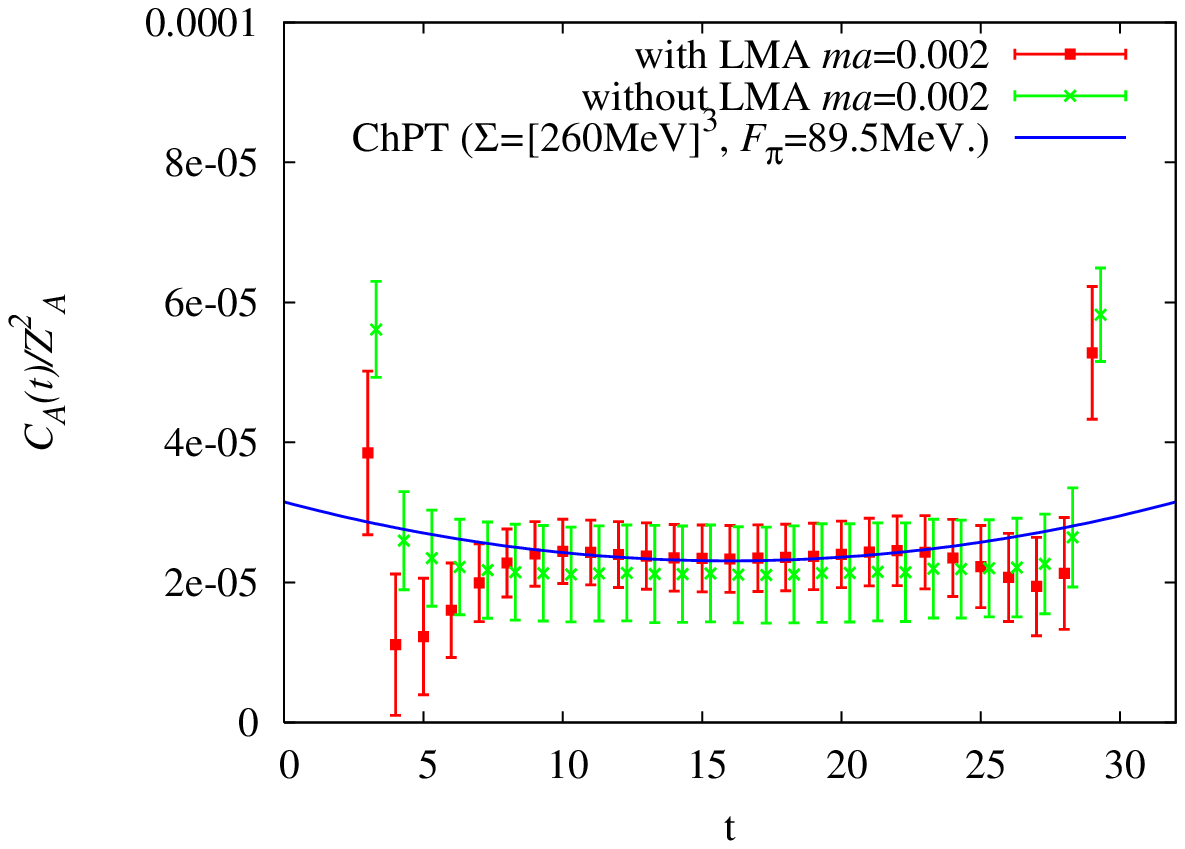}
\caption{Pseudoscalar and axial vector correlators computed by JLQCD with overlap Dirac operator, $N_{\rm f}=2$ \protect\cite{lat07:Fukaya}. }\label{fig:2p_jlqcd}
\vspace{-0.3cm}
\end{figure}

JLQCD presented at this conference preliminary unquenched results obtained with overlap Dirac operator and Iwasaki gauge action, 
$N_{\rm f}=2$ \cite{lat07:Fukaya}. They computed $PP$, $SS$, $V_0V_0$ and $A_0A_0$ correlators at a lattice spacing $a=0.11$ fm, $V=16^3\times 32$ and fixed topology $\nu=0$. Fig. \ref{fig:2p_jlqcd} shows the time-dependence of $PP$ and $A_0A_0$ correlators, for $am_{\rm sea}=am_{\rm val}=0.002$ corresponding to $\mu=0.556$. 
By simultaneously fitting pseudoscalar and axial correlators they obtain 
\begin{equation}
F=87.3(5.5) {\rm{MeV}},\;\;\;\;\;\;\;\Sigma^{\overline{\rm MS}}(2\;{\rm{GeV}})=[239.8(4.0){\rm{MeV}}]^3. 
\end{equation}
Pseudoscalar and current correlators have been recently computed in the effective theory at NLO also for non-degenerate quark masses in the full and partially quenched scenarios, in particular for the case where all quarks are in the $p$- or in the $\epsilon$-regime and for the mixed case, where $m_{\rm val}\Sigma V\lesssim 1$ and $m_{\rm sea}\Sigma V\gg 1$ \cite{Damgaard:2007ep,Bernardoni:2007hi}. These results may be very useful for the future matches of lattice QCD with the chiral effective theory.\\
An alternative method to extract pseudoscalar decay constant from two-point correlators in the $\epsilon$-regime has been proposed in \cite{Giusti:2003iq}: by matching 
residuals of $1/m^2$ poles with the expectations of the chiral effective theory, one can extract $F$ by means of zero-modes correlation functions at fixed non-zero topology. 
\subsection{LO LECs from eigenvalue distribution}
\begin{figure}
\vspace{-0.5cm}
\begin{center}
\includegraphics[width=7.3cm]{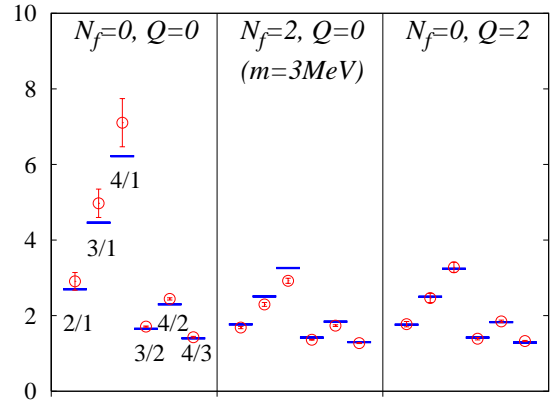}
\caption{Ratios of eigenvalues $\langle\zeta_k\rangle/ \langle\zeta_l\rangle$ for several $k,l$ computed in \protect\cite{Fukaya:2007yv}
compared with expectations of RMT.} \label{rmt_jlqcd}
\end{center}
\vspace{-0.3cm}
\end{figure}

\begin{figure}
\begin{center}
\includegraphics[width=7.3cm]{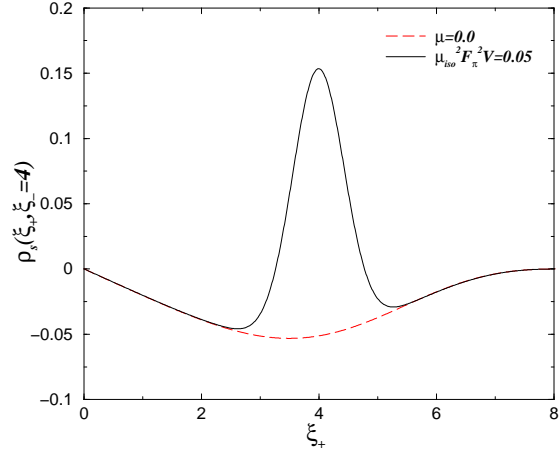}
\caption{The microscopic correlation function $\rho_{S,(1,1)}^{(1+1)})(\zeta_1,\zeta_2)$ in eq. \protect\ref{rho_mixed}
with $\mu_{\rm{iso},1}=-\mu_{\rm{iso},2}=  \mu_{\rm iso}$, $m_{u,d}\Sigma V =5$,  $\zeta_2=4$ ($\xi_-$ in the figure) , as a function of  $\zeta_1$ ($\xi_+$) \protect\cite{Damgaard:2006rh}. The red dashed curve represents the $ \mu_{\rm iso}=0$ case (with the delta-function omitted), while the black solid curve corresponds to $\mu_{\rm iso}^2 F^2V=0.05$.}\label{fig:iso}
\end{center}
\vspace{-0.5cm}
\end{figure}

\noindent
At LO in the $\epsilon$-expansion, the partition function at fixed topology is equivalent to the one of a chiral Random Matrix Theory (RMT) \cite{Shuryak:1992pi,Verbaarschot:1993pm,Verbaarschot:1994qf,Verbaarschot:2000dy}; it follows that RMT reproduces the same microscopic spectral density $\rho_{S\nu}(\zeta,\mu)$ as the chiral effective theory, in terms of  two dimensionless variables $\zeta=\lambda\Sigma V$ and $\mu=m\Sigma V$, where $\lambda$ are the eigenvalues of the Dirac operator. Moreover, it is possible to extract the probability distributions of single eigenvalues \cite{Nishigaki:1998is,Damgaard:2000ah,Basile:2007ki}: 
\begin{equation}
\rho_{S\nu}(\zeta,\mu)=\sum_{k=1}^\infty p_k^\nu(\zeta,\mu).
\end{equation}
By matching the low-lying spectrum with these expectations 
\begin{equation}
\langle\lambda_k\rangle_\nu^{\rm QCD}\Sigma V(\mu)
=\langle\zeta_k\rangle_\nu^{\rm RMT}(\mu)=\int d\zeta_k \zeta_kp_k^\nu(\zeta_k,\mu),\;\;\;\;\;\;(\lambda\ll F^2/\Sigma L^2)
\end{equation}
one can then extract the low-energy constant $\Sigma$.\\ 
The low-lying spectrum of the Dirac operator in the quenched approximation has been matched with the predictions of RMT in many works \cite{Edwards:1999ra,Bietenholz:2003mi,Giusti:2003gf,Wennekers:2005wa}. In \cite{Giusti:2003gf} a detailed comparison has been made, finding a good agreement for $L\gtrsim 1.5$ fm. Starting from this knowledge,
several unquenched computations have been performed in the past years \cite{DeGrand:2006nv,Lang:2006ab,Fukaya:2007yv,Hasenfratz:2007yj}
: their relevant parameters are summarised in table \ref{rmt_cond}.
JLQCD/TWQCD \cite{Fukaya:2007yv}  performed two separated sets of simulations, in the $\epsilon$- and the $p$-regime; in the last case they studied the sea quark mass dependence of $\Sigma$ and verified that its extrapolation to the chiral limit is compatible with the result obtained in the $\epsilon$-regime. Moreover, the sea quark mass dependence of ratios of eigenvalues for $N_{\rm f}=2, \nu=0$ is such that close to the massless limit reproduces the $N_{\rm f}=0, \nu=2$ results (\emph{flavor-topology duality}), while at heavy masses becomes compatible with $N_{\rm f}=0, \nu=0$.
Notice that in \cite{DeGrand:2006nv,Lang:2006ab} quark masses are still in a regime where $m\Sigma V$ is considerably larger than 1; nevertheless the observed quark-mass dependence of the extracted low-energy constant is negligible within the statistical errors. 
In figure \ref{rmt_jlqcd} the results of JLQCD/TWQCD are shown as an example. Ratios of eigenvalues 
  $\langle\zeta_k\rangle/ \langle\zeta_l\rangle$ are compared with the expectations of RMT; in particular 
the central plot corresponds to $N_{\rm f}=2,\nu=0,\mu=0.556(16)$; the left plot refers to $N_{\rm f}=0,\nu=0$ and the right to $N_{\rm f}=0,\nu=2$.
The physical results for the quark condensate in the $\overline{\rm MS}$ scheme at 2 GeV are reported in the last column of table \ref{rmt_cond}
\footnote{For \cite{Lang:2006ab,Hasenfratz:2007yj} the computation of the renormalisation constant $Z_S$ is still missing.}.\\
An important issue concerning these computations is the estimation of higher order corrections:
the matching with RMT is valid only at leading order and hence no control of higher order effects is possible. 
Assuming that finite-volume corrections for $\Sigma$ are the same as in the chiral effective theory, one would obtain a systematic error
of 31(11) MeV respectively on $\Sigma^{\overline{\rm MS}}(2 {\rm GeV})^{1/3}$ for \cite{DeGrand:2006nv}(\cite{Fukaya:2007yv}).
Another open question is that, while it is clear how the spectral density is renormalised \cite{DelDebbio:2005qa},
this is not the case for the individual eigenvalues.\\
\begin{table}
\begin{center}
\begin{tabular}{|c| c| c| c| c| c| c|c| }
\hline
& & & & & & & \\[-0.3cm]
{\emph{Authors}}  &   Dirac op.  & $N_{\rm f}$ & $a$ (fm) & $L/a,T/a$ & $m\Sigma V$  & $\nu$ & $\Sigma^{\overline{\rm MS}}(2\; {\rm GeV})$ \\
         & & & & & & & $({\rm MeV})^3$\\ 
\hline
\protect\cite{DeGrand:2006nv} & overlap & 2 & 0.15 & 10,10 &  $\simeq$ 2-5 & 0,1 & 282(10)\\
\hline
\protect\cite{Lang:2006ab}& chirally impr.   & 2  & 0.11-0.13 &  12,24  &  $\simeq$ 5-15 & 0,1 & \\
\hline
\protect\cite{Fukaya:2007yv}& overlap & 2 & 0.11 & 16,32 & 0.556(16)  & 0 &  251(7)\\
                            &         &   & 0.11-0.12 & 16,32 & $\simeq$ 7-30 & 0,2,4 & \\
\hline
\protect\cite{Hasenfratz:2007yj,lat07:Hierl} & fixed point  & 2+1 & 0.13 & 12,12 & $\simeq$ 1.5 (u,d), $\simeq$ 13 (s) & 0,1,2 &\\
\hline
\end{tabular}
\caption{Summary of simulation parameters of recent $N_{\rm f}=2$ simulations performed to extract the quark condensate by matching low-lying spectrum with RMT predictions.}\label{rmt_cond}
\end{center}
\end{table}
This framework can be extended such that the spectrum of the Dirac operator is sensitive also to the pseudoscalar decay constant $F$ at LO in the chiral effective theory; a proposal has been made in \cite{Damgaard:2005ys,Damgaard:2006pu,Damgaard:2006rh,Akemann:2006ru}.
One considers the eigenvalue equations of the Dirac operator in presence of external constant Abelian gauge potentials with couplings $\mu_{{\rm iso},1}\neq \mu_{{\rm iso},2}$:
\begin{equation}
D_{1,2}\psi_{1,2}^{(n)}  \equiv  \left[\Dslash(A)+i{\mu_{\rm{iso},1,2}}\gamma_0\right]\psi_{1,2}^{(n)}=i\lambda_{1,2}^{(n)}\psi_{1,2}^{(n)}.
\end{equation}
Translated into the chiral effective theory language, it leads to a modified partition function, which at LO in the $\epsilon$-regime reads:
\begin{equation}
\mathcal{Z}_\nu=\int_{\rm U(N)}dU_0 \left({\rm det}U_0  \right)^\nu \exp\left[\frac{1}{4}VF^2{\rm Tr}[U_0,B][U_0^\dagger,B]+\frac{1}{2}\Sigma V {\rm Tr}\left(\mathcal{M}^\dagger U_0+\mathcal{M}U_0^\dagger  \right)             \right],
\end{equation}
with 
\begin{equation}
B={\rm diag}\left(\mu_{{\rm iso},1} \mathbf{1}_{N_1},\mu_{{\rm iso},2}\mathbf{1}_{N_2}       \right);\;\;\;\; N_{\rm f}=N_1+N_2.
\end{equation}
Note that the imaginary nature of the chemical potential guarantees a positive definite fermion determinant.
Moreover, generalised Leutwyler-Smilga spectral sum rules can be derived \cite{Luz:2006vu}.\\
The same partition function can be obtained starting from a chiral Random Two-Matrix Theory \cite{Akemann:2006ru} (R2MT), from which it is possible to compute the spectral correlation functions in the microscopic limit $\rho_{S,(n,k)}^{(N_{\rm f})})(\{\zeta_1\}_n,\{\zeta_2 \}_k)$ for every $N_{\rm f}=N_1+N_2$ and $(n,k)$. The predictions from R2MT -or equivalently from the LO chiral Lagrangian-
are now formulated in terms of the rescaled variables $\zeta_1=\lambda_1\Sigma V$, $\zeta_2=\lambda_2\Sigma V$ and $\mu_f=m_f\Sigma V$. 
Moreover, the coupling with the isospin chemical potential introduces a new dependence on the combinations $\mu_{{\rm iso},1,2}F\sqrt{V}$, opening the possibility to extract $F$ by matching appropriate correlation functions measured on the lattice with the expectations of the LO effective theory. 
A typical example is the mixed correlation function
\begin{equation}\label{rho_mixed}
\rho_{(1,1)}^{(N_{\rm f})}(\lambda_1,\lambda_2)= \langle \sum_n \delta(\lambda_1-\lambda_1^{(n)}) \sum_m \delta(\lambda_2-\lambda_2^{(m)}) \rangle
-\langle \sum_n \delta(\lambda_1-\lambda_1^{(n)})   \rangle\langle \sum_m \delta(\lambda_2-\lambda_2^{(m)})          \rangle.
\end{equation}
The effect of having a non-zero chemical potential can be observed for instance in fig. \ref{fig:iso} \cite{Damgaard:2006rh}, where the prediction of the microscopic correlation function is plotted in the dynamical $N_{\rm f}=2$ case, with $N_1=N_2=1$, $\mu_{\rm{iso},1}=-\mu_{\rm{iso},2}=  \mu_{\rm iso}$. The variable $\zeta_2=4$ is fixed, and the $\zeta_1$-dependence is plotted: for $\mu_{\rm iso}=0$ the curve would show a delta-function in correspondence of $\zeta_1=4$; for $\mu_{\rm iso}\neq 0$ one observes a non-zero width. The quark masses in this case would change the height of the curve. 
As for the ordinary random matrix theory, it is theoretically possible to extract individual eigenvalue probability distributions in terms of all density correlators. Truncated expansions have been investigated and presented at this conference \cite{Akemann:2007wf}.\\
First numerical tests of the method have been performed in \cite{Damgaard:2005ys,Damgaard:2006pu}, using unimproved staggered quarks at coarse lattice spacing, both in the quenched and the dynamical $N_{\rm f}=2$ case.\\
A pilot study with overlap fermions, $N_{\rm f}=2$ \cite{DeGrand:2007tm}, has been presented at this conference \cite{lat07:DeGrand}; a lattice spacing $a\simeq 0.13$ fm has been adopted, with $L/a=T/a=12$ and $m\Sigma V\simeq 4$. In this work, the case $N_1=N_{\rm f}=2$, $N_2=0$, $\mu_{{\rm iso},1}=0$, $\mu_{{\rm iso},2}=\mu_{\rm iso}$ has been considered: it corresponds to a partially quenched situation, with two sea quarks at chemical potential $\mu_{{\rm iso},1}=0$ and two valence quarks coupled to $\mu_{\rm iso}$. The fit of the integrated correlator yields the results $\Sigma^{\overline{MS}}(2\;{\rm GeV})=[234(4)\;{\rm MeV}]^3$ (with finite-volume corrections estimated from chiral effective theory) and $F=101(6)$ MeV (with no finite-volume correction).\\
This method to extract $F$ can be relatively cheap compared for instance to the computation of current correlators; however, as already mentioned, NLO are not under control and -in absence of theoretical developments in this sense- simulations at several volumes would be needed in order to estimate finite-size effects.  

\vspace{-0.3cm}
\section{Conclusions}
\noindent
Lattice QCD simulations are now reaching quark masses and volumes where the matching with the chiral effective theory can be performed in a reliable way, 
keeping the different sources of systematic errors under control.
Many results are already available in the $p$-regime for $N_{\rm f}=2$; from the NLO quark-mass dependence of $M_\pi$ and $F_\pi$ it is possible to extract the LECs $\overline{l}_3$, $\overline{l}_4$, $F$ and $\Sigma$. Many efforts have been spent to evaluate discretisation errors and finite-volume corrections. Current simulations are now reaching pion masses of 300 MeV; in order to control uncertainties from higher order corrections in the chiral effective theory, lighter quark masses would be needed. With the recent algorithmic improvements and increasing computational resources, this is a realistic goal for the next years. \\
New results are also available from $N_{\rm f}=2+1$ simulations: besides the MILC results, RBC/UKQCD and PACS-CS presented preliminary studies at this conference, obtained with domain wall and $O(a)$ improved Wilson fermions respectively. From the quark-mass dependence of pseudoscalar masses and decay constant it will be possible to give reliable estimations of $F_0$, $\Sigma_0$, $L_4,L_5,L_6,L_8$.

In the $\epsilon$-regime, $\Sigma$ and $F$ can be extracted from observables which at NLO are not contaminated by higher order LECs. 
One can obtain fully independent determinations with respect to the ones achieved in the $p$-regime; their agreement is non-trivial test of QCD in the chiral regime. New unquenched results have been presented at this conference, both for the low-lying Dirac spectrum and meson correlation functions; although the $\epsilon$-regime simulations are challenging from the computational point of view, they provide an alternative and complementary approach to the standard $p$-regime determinations.

In this review only QCD LECs have been considered; however, lattice computations can be used for other constants which encode low-energy properties, for instance to evaluate couplings of the chiral weak Hamiltonian or of the Heavy Meson chiral effective theory.

\section{Acknowledgements}
I would like to thank P. Hern\'andez and L. Giusti for discussions and suggestions for the preparation of the talk and for the compilation of this manuscript.
Moreover I thank C. Haefeli, H. Leutwyler, P. Damgaard, G. Schierholz, P. Hasenfratz, C. Bernard, C. Urbach, S. Schaefer, T. DeGrand, J. Noaki, H. Fukaya, E. Scholz and A. Shindler for fruitful discussions and correspondence.




\begin{thebibliography}{10}

\bibitem{Weinberg:1978kz}
S.~Weinberg, {\em Physica} {\bf A96}
  (1979) 327.

\bibitem{Gasser:1983yg}
J.~Gasser and H.~Leutwyler, {\em
  Ann. Phys.} {\bf 158} (1984) 142.

\bibitem{Gasser:1984gg}
J.~Gasser and H.~Leutwyler, {\em Nucl. Phys.} {\bf B250} (1985) 465.

\bibitem{lat07:Bijnens}
J.~Bijnens, 
  {\pos{PoS(LATTICE 2007)004}} (2007),
  [\href{http://xxx.lanl.gov/abs/arXiv:0708.1377 [hep-lat]}{{\tt
  arXiv:0708.1377 [hep-lat]}}].

\bibitem{Gasser:1986vb}
J.~Gasser and H.~Leutwyler,   {\em Phys.
  Lett.} {\bf B184} (1987) 83.

\bibitem{Gasser:1987ah}
J.~Gasser and H.~Leutwyler,  {\em
  Phys. Lett.} {\bf B188} (1987) 477.

\bibitem{Leutwyler:1992yt}
H.~Leutwyler and A.~Smilga,   {\em Phys. Rev.} {\bf D46} (1992) 5607--5632.

\bibitem{Bijnens:1996wm}
J.~Bijnens and P.~Talavera, 
  {\em Nucl. Phys.} {\bf B489} (1997) 387--404,
  [\href{http://xxx.lanl.gov/abs/hep-ph/9610269}{{\tt hep-ph/9610269}}].

\bibitem{Colangelo:2001df}
G.~Colangelo, J.~Gasser, and H.~Leutwyler, {\em Nucl.
  Phys.} {\bf B603} (2001) 125--179,
  [\href{http://xxx.lanl.gov/abs/hep-ph/0103088}{{\tt hep-ph/0103088}}].

\bibitem{Colangelo:2003hf}
G.~Colangelo and S.~D\"urr,  {\em Eur.
  Phys. J.} {\bf C33} (2004) 543--553,
  [\href{http://xxx.lanl.gov/abs/hep-lat/0311023}{{\tt hep-lat/0311023}}].

\bibitem{DelDebbio:2006cn}
L.~Del~Debbio, L.~Giusti, M.~Luscher, R.~Petronzio, and N.~Tantalo, {\em JHEP} {\bf 02} (2007) 056,
  [\href{http://xxx.lanl.gov/abs/hep-lat/0610059}{{\tt hep-lat/0610059}}].

\bibitem{Luscher:2003qa}
M.~L\"uscher,  {\em Comput. Phys. Commun.} {\bf 156} (2004)
  209--220, [\href{http://xxx.lanl.gov/abs/hep-lat/0310048}{{\tt
  hep-lat/0310048}}].

\bibitem{Luscher:2005rx}
M.~L\"uscher,  {\bf 165} (2005) 199--220,
  [\href{http://xxx.lanl.gov/abs/hep-lat/0409106}{{\tt hep-lat/0409106}}].

\bibitem{Boucaud:2007uk}
{\bf ETM} Collaboration, P.~Boucaud {\em et~al.}, {\em Phys. Lett.} {\bf B650} (2007) 304--311,
  [\href{http://xxx.lanl.gov/abs/hep-lat/0701012}{{\tt hep-lat/0701012}}].

\bibitem{lat07:Urbach}
C.~Urbach, 
  {\pos{PoS(LATTICE 2007)022}}, (2007) [\href{http://xxx.lanl.gov/abs/arXiv:0710.1517 [hep-lat]}{{\tt
  arXiv:0710.1517 [hep-lat]}}].


\bibitem{lat07:Herdoiza}
G.~Herdoiza,   {\pos{PoS(LATTICE 2007)102}}.

\bibitem{Colangelo:2005gd}
G.~Colangelo, S.~Durr, and C.~Haefeli,  {\em Nucl. Phys.} {\bf B721} (2005) 136--174,
  [\href{http://xxx.lanl.gov/abs/hep-lat/0503014}{{\tt hep-lat/0503014}}].

\bibitem{lat07:Schierholz}
G.~Schierholz,   {\pos{PoS(LATTICE 2007)133}}.

\bibitem{lat07:Matsufuru}
H.~Matsufuru,  {\pos{PoS(LATTICE 2007)018}}.

\bibitem{lat07:Noaki}
J.~Noaki {\em et~al.},   { \pos{PoS(LATTICE 2007)126}} (2007),
  [\href{http://xxx.lanl.gov/abs/arXiv:0710.0929 [hep-lat]}{{\tt
  arXiv:0710.0929 [hep-lat]}}].

\bibitem{Aoki:2007ka}
S.~Aoki, H.~Fukaya, S.~Hashimoto, and T.~Onogi,   \href{http://xxx.lanl.gov/abs/arXiv:0707.0396
  [hep-lat]}{{\tt arXiv:0707.0396 [hep-lat]}}.

\bibitem{Sommer:1993ce}
R.~Sommer,
   {\em Nucl. Phys.} {\bf B411} (1994) 839--854,
  [\href{http://xxx.lanl.gov/abs/hep-lat/9310022}{{\tt hep-lat/9310022}}].

\bibitem{Leutwyler:2007ae}
H.~Leutwyler,
  \href{http://xxx.lanl.gov/abs/arXiv:0706.3138 [hep-ph]}{{\tt arXiv:0706.3138
  [hep-ph]}}.

\bibitem{Sharpe:2000bc}
S.~R. Sharpe and N.~Shoresh, {\em Phys. Rev.} {\bf D62} (2000) 094503,
  [\href{http://xxx.lanl.gov/abs/hep-lat/0006017}{{\tt hep-lat/0006017}}].

\bibitem{Bijnens:2006jv}
J.~Bijnens, N.~Danielsson, and T.~A. Lahde,  {\em Phys. Rev.} {\bf D73} (2006) 074509,
  [\href{http://xxx.lanl.gov/abs/hep-lat/0602003}{{\tt hep-lat/0602003}}].

\bibitem{Sharpe:2006pu}
S.~R. Sharpe,
  \href{http://xxx.lanl.gov/abs/hep-lat/0607016}{{\tt hep-lat/0607016}}.

\bibitem{Aubin:2004fs}
{\bf MILC} Collaboration, C.~Aubin {\em et~al.}, {\em Phys. Rev.} {\bf D70} (2004) 114501,
  [\href{http://xxx.lanl.gov/abs/hep-lat/0407028}{{\tt hep-lat/0407028}}].

\bibitem{Bernard:2006wx}
{\bf MILC} Collaboration, C.~Bernard {\em et~al.},  {\em PoS} {\bf LAT2006} (2006) 163,
  [\href{http://xxx.lanl.gov/abs/hep-lat/0609053}{{\tt hep-lat/0609053}}].

\bibitem{lat07:Bernard}
C.~Bernard {\em et~al.}, {\pos{PoS(LATTICE 2007)090}} (2007),
  [\href{http://xxx.lanl.gov/abs/arXiv:0710.1118 [hep-lat]}{{\tt
  arXiv:0710.1118 [hep-lat]}}].

\bibitem{Lee:1999zxa}
W.-J. Lee and S.~R. Sharpe,   {\em Phys. Rev.} {\bf D60} (1999) 114503,
  [\href{http://xxx.lanl.gov/abs/hep-lat/9905023}{{\tt hep-lat/9905023}}].

\bibitem{Allton:2007hx}
{\bf RBC and UKQCD} Collaboration, C.~Allton {\em et~al.},  {\em Phys. Rev.} {\bf D76} (2007) 014504,
  [\href{http://xxx.lanl.gov/abs/hep-lat/0701013}{{\tt hep-lat/0701013}}].

\bibitem{lat07:Boyle}
P.~Boyle,  {\pos{PoS(LATTICE 2007)005}}.

\bibitem{lat07:Lin}
{\bf UKQCD} Collaboration, R.-. M.~F. Lin and E.~E. Scholz,  {\pos{PoS(LATTICE
  2007)120}} (2007), [\href{http://xxx.lanl.gov/abs/arXiv:0710.0536
  [hep-lat]}{{\tt arXiv:0710.0536 [hep-lat]}}].

\bibitem{lat07:Kuramashi}
Y.~Kuramashi, {\pos{PoS(LATTICE 2007)017}}.

\bibitem{lat07:Ukita}
N.~Ukita,  {\pos{PoS(LATTICE 2007)138}}.

\bibitem{DelDebbio:2005qa}
L.~Del~Debbio, L.~Giusti, M.~L\"uscher, R.~Petronzio, and N.~Tantalo, {\em
  JHEP} {\bf 02} (2006) 011,
  [\href{http://xxx.lanl.gov/abs/hep-lat/0512021}{{\tt hep-lat/0512021}}].

\bibitem{DelDebbio:2007pz}
L.~Del~Debbio, L.~Giusti, M.~L\"uscher, R.~Petronzio, and N.~Tantalo, {\em JHEP} {\bf 02} (2007) 082,
  [\href{http://xxx.lanl.gov/abs/hep-lat/0701009}{{\tt hep-lat/0701009}}].

\bibitem{lat07:Shindler}
A.~Shindler,
  {\pos{PoS(LATTICE 2007)084}}.

\bibitem{DeGrand:2004qw}
T.~A. DeGrand and S.~Schaefer, {\em Comput. Phys. Commun.} {\bf 159} (2004) 185--191,
  [\href{http://xxx.lanl.gov/abs/hep-lat/0401011}{{\tt hep-lat/0401011}}].

\bibitem{Giusti:2004yp}
L.~Giusti, P.~Hern\'andez, M.~Laine, P.~Weisz, and H.~Wittig, {\em JHEP}
  {\bf 04} (2004) 013, [\href{http://xxx.lanl.gov/abs/hep-lat/0402002}{{\tt
  hep-lat/0402002}}].

\bibitem{Neuberger:1987fd}
H.~Neuberger,   {\em Nucl. Phys.} {\bf B300}
  (1988) 180.

\bibitem{Neuberger:1987zz}
H.~Neuberger,  {\em Phys. Rev. Lett.} {\bf 60} (1988) 889.

\bibitem{Hasenfratz:1989pk}
P.~Hasenfratz and H.~Leutwyler, {\em
  Nucl. Phys.} {\bf B343} (1990) 241--284.

\bibitem{Hernandez:1999cu}
P.~Hern\'andez, K.~Jansen, and L.~Lellouch, {\em Phys. Lett.} {\bf B469} (1999)
  198--204, [\href{http://xxx.lanl.gov/abs/hep-lat/9907022}{{\tt
  hep-lat/9907022}}].

\bibitem{DeGrand:2001ie}
{\bf MILC} Collaboration, T.~A. DeGrand, {\em Phys. Rev.} {\bf D64} (2001)
  117501, [\href{http://xxx.lanl.gov/abs/hep-lat/0107014}{{\tt
  hep-lat/0107014}}].

\bibitem{Hasenfratz:2002rp}
P.~Hasenfratz, S.~Hauswirth, T.~Jorg, F.~Niedermayer, and K.~Holland, 
  {\em Nucl. Phys.} {\bf B643} (2002) 280--320,
  [\href{http://xxx.lanl.gov/abs/hep-lat/0205010}{{\tt hep-lat/0205010}}].

\bibitem{Giusticond}
L.~Giusti and S.~Necco, {\em JHEP} {\bf 04} (2007) 090,
  [\href{http://xxx.lanl.gov/abs/hep-lat/0702013}{{\tt hep-lat/0702013}}].

\bibitem{Hansen:1990un}
F.~C. Hansen,  {\em Nucl. Phys.} {\bf B345} (1990) 685--708.

\bibitem{Hansen:1990yg}
F.~C. Hansen and H.~Leutwyler,  {\em Nucl. Phys.} {\bf B350} (1991) 201--227.

\bibitem{Damgaard:2002qe}
P.~H. Damgaard, P.~Hern\'andez, K.~Jansen, M.~Laine, and L.~Lellouch, {\em Nucl.
  Phys.} {\bf B656} (2003) 226--238,
  [\href{http://xxx.lanl.gov/abs/hep-lat/0211020}{{\tt hep-lat/0211020}}].

\bibitem{Hernandez:2002ds}
P.~Hern\'andez and M.~Laine, {\em JHEP} {\bf 01} (2003)
  063, [\href{http://xxx.lanl.gov/abs/hep-lat/0212014}{{\tt hep-lat/0212014}}].

\bibitem{Hernandez:2006kz}
P.~Hern\'andez and M.~Laine, {\em JHEP} {\bf 10} (2006) 069,
  [\href{http://xxx.lanl.gov/abs/hep-lat/0607027}{{\tt hep-lat/0607027}}].

\bibitem{Bietenholz:2003bj}
W.~Bietenholz, T.~Chiarappa, K.~Jansen, K.~I. Nagai, and S.~Shcheredin, {\em JHEP} {\bf 02} (2004) 023,
  [\href{http://xxx.lanl.gov/abs/hep-lat/0311012}{{\tt hep-lat/0311012}}].

\bibitem{Fukaya:2005yg}
H.~Fukaya, S.~Hashimoto, and K.~Ogawa, {\em Prog. Theor. Phys.}
  {\bf 114} (2005) 451--476,
  [\href{http://xxx.lanl.gov/abs/hep-lat/0504018}{{\tt hep-lat/0504018}}].

\bibitem{lat07:Fukaya}
H.~Fukaya, {\pos{PoS(LATTICE 2007)073}}.

\bibitem{Damgaard:2007ep}
P.~H. Damgaard and H.~Fukaya, \href{http://xxx.lanl.gov/abs/arXiv:0707.3740
  [hep-lat]}{{\tt arXiv:0707.3740 [hep-lat]}}.

\bibitem{Bernardoni:2007hi}
F.~Bernardoni and P.~Hern\'andez,
  \href{http://xxx.lanl.gov/abs/arXiv:0707.3887 [hep-lat]}{{\tt arXiv:0707.3887
  [hep-lat]}}.

\bibitem{Giusti:2003iq}
L.~Giusti, P.~Hern\'andez, M.~Laine, P.~Weisz, and H.~Wittig, {\em JHEP} {\bf
  01} (2004) 003, [\href{http://xxx.lanl.gov/abs/hep-lat/0312012}{{\tt
  hep-lat/0312012}}].

\bibitem{Shuryak:1992pi}
E.~V. Shuryak and J.~J.~M. Verbaarschot,  {\em Nucl. Phys.} {\bf A560}
  (1993) 306--320, [\href{http://xxx.lanl.gov/abs/hep-th/9212088}{{\tt
  hep-th/9212088}}].

\bibitem{Verbaarschot:1993pm}
J.~J.~M. Verbaarschot and I.~Zahed,  {\em Phys. Rev. Lett.} {\bf 70} (1993)
  3852--3855, [\href{http://xxx.lanl.gov/abs/hep-th/9303012}{{\tt
  hep-th/9303012}}].

\bibitem{Verbaarschot:1994qf}
J.~J.~M. Verbaarschot, {\em Phys. Rev. Lett.} {\bf 72}
  (1994) 2531--2533, [\href{http://xxx.lanl.gov/abs/hep-th/9401059}{{\tt
  hep-th/9401059}}].

\bibitem{Verbaarschot:2000dy}
J.~J.~M. Verbaarschot and T.~Wettig,  {\em Ann. Rev. Nucl. Part. Sci.} {\bf 50} (2000) 343--410,
  [\href{http://xxx.lanl.gov/abs/hep-ph/0003017}{{\tt hep-ph/0003017}}].

\bibitem{Nishigaki:1998is}
S.~M. Nishigaki, P.~H. Damgaard, and T.~Wettig, {\em Phys. Rev.} {\bf D58} (1998)
  087704, [\href{http://xxx.lanl.gov/abs/hep-th/9803007}{{\tt
  hep-th/9803007}}].

\bibitem{Damgaard:2000ah}
P.~H. Damgaard and S.~M. Nishigaki,  {\em Phys. Rev.} {\bf D63} (2001) 045012,
  [\href{http://xxx.lanl.gov/abs/hep-th/0006111}{{\tt hep-th/0006111}}].

\bibitem{Basile:2007ki}
F.~Basile and G.~Akemann, 
  \href{http://xxx.lanl.gov/abs/arXiv:0710.0376 [hep-th]}{{\tt arXiv:0710.0376
  [hep-th]}}.

\bibitem{Edwards:1999ra}
R.~G. Edwards, U.~M. Heller, J.~E. Kiskis, and R.~Narayanan, {\em Phys. Rev. Lett.} {\bf 82}
  (1999) 4188--4191, [\href{http://xxx.lanl.gov/abs/hep-th/9902117}{{\tt
  hep-th/9902117}}].

\bibitem{Bietenholz:2003mi}
W.~Bietenholz, K.~Jansen, and S.~Shcheredin, {\em JHEP} {\bf 07} (2003) 033,
  [\href{http://xxx.lanl.gov/abs/hep-lat/0306022}{{\tt hep-lat/0306022}}].

\bibitem{Giusti:2003gf}
L.~Giusti, M.~L{\"u}scher, P.~Weisz, and H.~Wittig,   {\em JHEP} {\bf 11} (2003) 023,
  [\href{http://xxx.lanl.gov/abs/hep-lat/0309189}{{\tt hep-lat/0309189}}].

\bibitem{Wennekers:2005wa}
J.~Wennekers and H.~Wittig,    {\em JHEP} {\bf 09} (2005) 059,
  [\href{http://xxx.lanl.gov/abs/hep-lat/0507026}{{\tt hep-lat/0507026}}].

\bibitem{DeGrand:2006nv}
T.~DeGrand, Z.~Liu, and S.~Schaefer, 
  {\em Phys. Rev.} {\bf D74} (2006) 094504,
  [\href{http://xxx.lanl.gov/abs/hep-lat/0608019}{{\tt hep-lat/0608019}}].

\bibitem{Lang:2006ab}
C.~B. Lang, P.~Majumdar, and W.~Ortner, {\em Phys. Lett.} {\bf B649} (2007)
  225--229, [\href{http://xxx.lanl.gov/abs/hep-lat/0611010}{{\tt
  hep-lat/0611010}}].

\bibitem{Fukaya:2007yv}
H.~Fukaya {\em et~al.},   \href{http://xxx.lanl.gov/abs/arXiv:0705.3322
  [hep-lat]}{{\tt arXiv:0705.3322 [hep-lat]}}.

\bibitem{Hasenfratz:2007yj}
P.~Hasenfratz {\em et~al.}, 
  \href{http://xxx.lanl.gov/abs/arXiv:0707.0071 [hep-lat]}{{\tt arXiv:0707.0071
  [hep-lat]}}.

\bibitem{lat07:Hierl}
P.~Hasenfratz {\em et~al.}, {\pos{PoS(LATTICE 2007)077}} (2007),
  [\href{http://xxx.lanl.gov/abs/arXiv:0710.0551 [hep-lat]}{{\tt
  arXiv:0710.0551 [hep-lat]}}].

\bibitem{Damgaard:2005ys}
P.~H. Damgaard, U.~M. Heller, K.~Splittorff, and B.~Svetitsky,   {\em Phys. Rev.} {\bf D72}
  (2005) 091501, [\href{http://xxx.lanl.gov/abs/hep-lat/0508029}{{\tt
  hep-lat/0508029}}].

\bibitem{Damgaard:2006pu}
P.~H. Damgaard, U.~M. Heller, K.~Splittorff, B.~Svetitsky, and D.~Toublan, {\em Phys. Rev.}
  {\bf D73} (2006) 074023, [\href{http://xxx.lanl.gov/abs/hep-lat/0602030}{{\tt
  hep-lat/0602030}}].

\bibitem{Damgaard:2006rh}
P.~H. Damgaard, U.~M. Heller, K.~Splittorff, B.~Svetitsky, and D.~Toublan, {\em Phys. Rev.} {\bf D73} (2006) 105016,
  [\href{http://xxx.lanl.gov/abs/hep-th/0604054}{{\tt hep-th/0604054}}].

\bibitem{Akemann:2006ru}
G.~Akemann, P.~H. Damgaard, J.~C. Osborn, and K.~Splittorff, {\em
  Nucl. Phys.} {\bf B766} (2007) 34--67,
  [\href{http://xxx.lanl.gov/abs/hep-th/0609059}{{\tt hep-th/0609059}}].

\bibitem{Luz:2006vu}
M.~Luz, {\it Determining f(pi) from spectral sum rules},  {\em Phys. Lett.}
  {\bf B643} (2006) 235--239,
  [\href{http://xxx.lanl.gov/abs/hep-lat/0607022}{{\tt hep-lat/0607022}}].

\bibitem{Akemann:2007wf}
G.~Akemann and P.~H. Damgaard, {\pos{PoS(LATTICE2007)166}} (2007), [\href{http://xxx.lanl.gov/abs/arXiv:0709.0484
  [hep-lat]}{{\tt arXiv:0709.0484 [hep-lat]}}].

\bibitem{DeGrand:2007tm}
T.~DeGrand and S.~Schaefer,
  \href{http://xxx.lanl.gov/abs/arXiv:0708.1731 [hep-lat]}{{\tt arXiv:0708.1731
  [hep-lat]}}.

\bibitem{lat07:DeGrand}
T.~DeGrand and S.~Schaefer, {\pos{PoS(LATTICE 2007)069}} (2007),
  [\href{http://xxx.lanl.gov/abs/arXiv:0709.2889 [hep-lat]}{{\tt
  arXiv:0709.2889 [hep-lat]}}].

\end{thebibliography}

\providecommand{\href}[2]{#2}\begingroup\raggedright\endgroup

\end{document}